\documentclass[%
 aip,
 rsi,
 amsmath,amssymb,
reprint,%
]{revtex4-1}

\usepackage{graphicx}
\usepackage{dcolumn}
\usepackage{bm}

\usepackage[utf8]{inputenc}
\usepackage[T1]{fontenc}
\usepackage{mathptmx}

\begin{document}

\preprint{AIP/123-QED}

\title{Set-up for multi coincidence experiments of EUV to VIS photons and charged particles – the solid angle maximization approach}
\author{A. Hans}
\email[Equally contributing author. Email: ]{hans@physik.uni-kassel.de}
\author{C. Ozga}
\email[Equally contributing author. Email: ]{ozga@physik.uni-kassel.de}
\author{Ph. Schmidt}%
\author{G. Hartmann}%
\author{A. Nehls}%
\author{Ph. Wenzel}%
\affiliation{ 
Institute of Physics and Center for Interdisciplinary Nanostructure Science and Technology (CINSaT), University of Kassel, Heinrich-Plett-Straße 40, 34132 Kassel, Germany
}%
\author{C. Richter}
\affiliation{ 
Leibniz Institute of Surface Engineering (IOM), Permoserstr. 15, 04318 Leipzig, Germany
}
\author{C. Lant}
\affiliation{ 
Department of Physics, New York University, 726 Broadway, New York 10003, USA
}
\author{\newline X. Holzapfel}
\author{J.H. Viehmann}
\affiliation{ 
Institute of Physics and Center for Interdisciplinary Nanostructure Science and Technology (CINSaT), University of Kassel, Heinrich-Plett-Straße 40, 34132 Kassel, Germany
}%
\author{U. Hergenhahn}
\thanks{ 
also at: Max Planck Institute for Plasma Physics, Wendelsteinstr. 1, 17491 Greifswald, Germany; Fritz-Haber-Institut der Max-Planck-Gesellschaft, Faradayweg 4-6, 14195 Berlin, Germany
}
\affiliation{ 
Leibniz Institute of Surface Engineering (IOM), Permoserstr. 15, 04318 Leipzig, Germany
}
\author{A. Ehresmann}
\author{A. Knie}
\affiliation{ 
Institute of Physics and Center for Interdisciplinary Nanostructure Science and Technology (CINSaT), University of Kassel, Heinrich-Plett-Straße 40, 34132 Kassel, Germany
}%

\date{\today}

\begin{abstract}
The coincident detection of particles is a powerful method in experimental physics, enabling the investigation of a variety of projectile-target interactions.
The vast majority of coincidence experiments is performed with charged particles, as they can be guided by electric or magnetic fields to yield large detection probabilities.
When a neutral species or a photon is one of the particles recorded in coincidence, its detection probability typically suffers from small solid angles.
Here, we present two optical assemblies considerably enhancing the solid angle for EUV to VIS photon detection.
The efficiency and versatility of these assemblies is demonstrated for electron-photon coincidence detection, where electrons and photons emerge from fundamental processes after photoexcitation of gaseous samples by synchrotron radiation.\footnote{The following article has been submitted to \emph{Review of Scientific Instruments} (AIP). After it is published, it will be found at \url{https://aip.scitation.org/journal/rsi}.}
\end{abstract}

\maketitle


\section{\label{sec1}Introduction}
In atomic, molecular, or cluster physics, the interaction of a target system with a projectile typically leads to its excitation, ionization, or in case of bound systems to their fragmentation with corresponding emissions of a variety of particles such as electrons, ions, neutral species and photons.
For processes that produce excited states of the target, there is in general not a singular de-excitation path, but branching into competing channels.
The weaker processes are often difficult to identify and even harder to quantify, as their occurrence may be masked by other, more intense signals or they may barely emerge above the noise level at all.\\
The simultaneous detection of several or all reaction products belonging to a particular projectile-target interaction allows to disentangle individual decay pathways and to increase the signal to noise contrast for these weak processes significantly.
This powerful technique is called coincidence measurement and is widely used in experimental particle, nuclear, atomic, molecular, and cluster physics (Ref.~\onlinecite{Arion2015} and references therein).
The most essential experimental parameters in a coincidence experiment are the detection probabilities $P_{i} (\in [0,1])$ of the respective particles $i$, which may be separated into products of the accepted relative solid angles $\Omega_{\text{rel},i} (\in [0,1])$ and the detector efficiencies $\varepsilon_{i} (\in [0,1])$.
The total probability of recording a coincident event $P_\text{coinc}$ is the product of the individual detection probabilities of all involved particles:
\begin{eqnarray}
P_\text{coinc}=\prod\limits_{\text{all particles}}{P_i}=\prod\limits_\text{all particles}{\Omega_{\text{rel},i}\varepsilon_i}
\label{pcoinc_eq}.
\end{eqnarray}
The detector efficiency $\varepsilon_i$ is the probability that a particle reaching the detector is recorded as an actual event.
It is an intrinsic characteristic of the detector and typically depends on the particle type and its properties such as energy and mass.
The achievable solid angle, however, is given by the experimental geometry.
In general, it can be described by the active detector surface area $A_\text{D}$ and its distance to the interaction volume $D$:
\begin{eqnarray}
\Omega_\text{rel}\approx\frac{A_\text{D}}{4\pi D^2}
\label{photodet_eq}.
\end{eqnarray}
In the case of charged reaction products (electrons or ions), $\Omega_\text{rel}$ often can be increased to values close to unity by guiding these particles towards the active detector area via electric and/or magnetic fields.
A prominent example for the latter case are “reaction microscope” experiments in which charged fragments of atoms, molecules, or clusters are measured in multiple coincidences  \cite{Doerner2000,Ullrich2003}.
However, since guiding fields are not applicable to neutral fragments and photons, the solid angle for the detection of these particles is typically small and results in a low coincidence rate.
Because of the above mentioned challenge, reports of successful coincidence measurements with detection of these kind of particles are rare compared to ones restricted to charged particles.
In the following we will restrict the discussion to experiments in which a charged particle and a photon from the same ionization event are recorded.\\
One class of these experiments uses excitation by monochromatized emission of noble gas lamps to investigate fluorescing ionic fragments of various molecular systems ranging from diatomic N$_2$ or CO to fluorobenzene molecules \cite{Bloch1975,Eland1976,Dujardin1981,Field1992}.
Another approach was the measurement of scattering angles in  electron energy loss spectroscopy in coincidence with an emitted photon to investigate doubly-excited states in H$_2$ and H$_2$O \cite{Ishikawa2011,Tsuchida2011} as well as interferences in the He(3l,3l’) excitation \cite{Dogan1998}.\\
However, these measurements do not use the opportunities offered by modern synchrotron radiation facilities, in particular the narrow bandwidth of the exciting radiation in combination with the high tunability of the exciting-photon energies.
This type of excitation was used for photon-threshold electron measurements which allow to determine the lifetime of radiative molecular states\cite{Schlag1977}.
Photoion-photon coincidence techniques were also applied at synchrotron radiation facilities to investigate for example the dissociative photoionization of N$_2$ \cite{Kitajima1995,Kitajima1996,Soderstroma2004} and the core-hole decay of argon atoms and clusters \cite{Gejo2013}.\\
A series of experiments showed that the coincident measurement of photons and electrons, combined with polarization analysis of the fluorescence, for atomic photoionization can be used to perform so-called complete experiments, in which all amplitudes and phases in a partial wave description of the process are determined. \cite{Beyer1995,Beyer1996,West1998,Arp1996}.
Yet, all these studies suffer from low coincidence rates, compensated to some extent by using long acquisition times \cite{Bloch1975}.
However, if the experiment needs to make use of synchrotron radiation, for which allocated beamtime is limited, reasonable statistics might not be achievable or losses in data quality have to be accepted.\\
These challanges can be overcome according to the relation given in equation \ref{pcoinc_eq}, if collecting optics are used to increase the solid angle of photon detection and with that increase the coincidence count rate as described in Ref.~\onlinecite{Reiss2015} for photon-photon and in Ref.~\onlinecite{Meyer1995} for ion-photon coincidence experiments.\\
In this paper, we show how to improve the usability of optical assemblies dedicated to the coincident detection of a photon and a charged particle.
These assemblies dramatically enhance the solid angle of photon detection and enable efficient coincidence experiments with at least one participating photon.
We present two configurations:
A) A mirror assembly surrounding the interaction volume.
This design is optimized for maximum solid angle.
B) A combination of flexible optical elements adaptable to a variety of experimental constraints, suited for experiments in which a direct view on the interaction volume is impossible. 
The applicability of both designs is demonstrated by performing electron-photon coincidence measurements on fundamental processes after excitation of supersonic noble gas jets by synchrotron radiation.

\section{\label{sec2}Coincident detection of electrons and photons}
\begin{figure}
\includegraphics{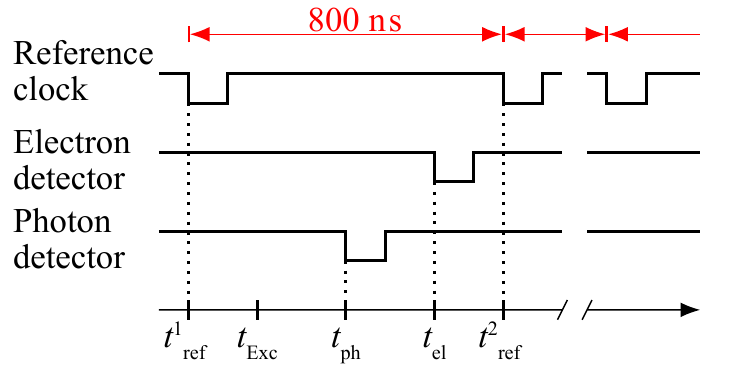}
\caption{\label{fig_detscheme}Detection scheme of electron-photon coincidences.
The TDC is triggered by the reference clock and records the arrival times of the electrons $t_\text{el}$ and photons $t_\text{ph}$ relative to the reference clock pulse $t_\text{ref}$.
At synchrotron radiation facilities, the time of excitation $t_\text{exc}$ typically has a constant offset to the provided reference clock. For the showcase experiments, the time of $800\,\text{ns}$ between two excitations corresponds to the circulation time of an electron bunch in BESSY II in single bunch mode.}
\end{figure}

In this section, we explain the timing scheme of our experiment, followed by other details of the set-up.
In many coincidence experiments, including the ones described in this work, either pulsed target delivery or pulsed excitation sources like pulsed lasers, ion bunches in ion storage facilities, or synchrotron radiation pulses with an appropriate reference clock for the measurements are used to ensure that all measured particles have originated in the same physical event.
The reference clock pulse is used as a start signal for a time-to-digital converter (TDC).
In the case of electron-photon coincidence experiments, a coincidence event occurs, if at least one electron and one photon are detected after a common start signal.
In general, coincidence events can result from into true or accidental (or false\cite{Wehlitz1995}) coincidences.
For true coincidences, both detected particles originate in the same interaction process.
Accidental coincidences occur, if the electron and the photon originate from two different physical processes or from two independent interactions at different sites of the sample.
For simplicity and without loss of generality, it is assumed in this work that for a true coincidence event, both the photon and the electron reach the respective detector within the time interval between two consecutive excitation pulses.\\
Often, true and accidental coincidences bear no experimental signature that allows their separation on an event-by-event basis.
A method to eliminate accidental coincidences uses data acquisition over several consecutive excitation pulses.
It is assumed that the rate of accidental coincidences is the same when both particles are detected coincidentally between one reference clock pulse and the successive one as compared to the accidental coincidence signal when the two particles are recorded individually with respect to different successive reference clock pulses.
Therefore, the true coincidence spectrum can be obtained by the subtraction of the pure accidental coincidence spectrum from the total coincidence spectrum.
This is a purely statistical method and does not allow identification of individual true coincidences.\\  
In a typical time of flight spectrometer, the electron spectrum is obtained by collecting the arrival times of all electron events in a histogram and the time axis encodes the kinetic energy, while the respective histogram of the photon events yields the information about the lifetime of the radiative state.
The signal acquisition of an electron-photon-coincidence event is shown in Figure \ref{fig_detscheme}.
The true coincident electron spectrum is explicitly obtained by separation of the total coincident electron spectrum into four cases:
\begin{itemize}
\item Both electron and photon were detected between the first and the second reference clock pulse ($t_\text{el}, t_\text{ph}<t^2_\text{ref}$). These events are named $\text{sp}_{11}$. 
\item The electron was detected between the first and the second pulse of the reference clock and the photon was detected after the second reference clock signal ($t_\text{el}<t^2_\text{ref}$ and $t_\text{ph}>t^2_\text{ref}$), named $\text{sp}_{12}$.
\item The electron was detected after the second reference clock pulse and the photon was detected between the first and the second reference clock signal ($t_\text{el}>t^2_\text{ref}$ and $t_\text{ph}<t^2_\text{ref}$). These events are named $\text{sp}_{21}$.
\item Both electron and photon were detected after the second reference clock pulse ($t_\text{el},t_\text{ph}>t^2_\text{ref}$), named $\text{sp}_{22}$.
\end{itemize}
The spectrum of true coincidences is then obtained by:
\begin{eqnarray}
\text{True coincidences}=\text{sp}_{11}+\text{sp}_{22}-\text{sp}_{12}-\text{sp}_{21}
\label{coinc_eq}
\end{eqnarray}
This method of data acquisition and processing also allows the detection of multiple electrons with a photon in coincidence.\\
Proof-of-principle experiments were performed at the BESSY~II storage ring of the Helmholtz-Zentrum Berlin.
In all examples, the synchrotron was operated in single bunch mode with a circulation time of $800\,\text{ns}$, i.e. $800\,\text{ns}$ temporal spacing between two subsequent excitations.
In the presented examples, magnetic bottle type electron spectrometers were used for electron detection.
Details for one of the instruments can be found in Ref.~\onlinecite{Mucke2012}.
Briefly, the spectrometer uses the inhomogeneous field of a magnetic tip to collect and redirect the electrons from the interaction volume towards a drift tube.
A magnetic field parallel to the tube axis prevents the loss of electrons due to lateral velocity components. At the end of the drift tube, electrons are amplified by a chevron stack of microchannel plates (MCPs) \cite{Wiza1979}.
Detection is carried out by an anode, where the corresponding voltage drop is retrieved from the high voltage potential using capacitive coupling \cite{Mucke2012}.
The signal is processed by a constant fraction discriminator and its arrival time relative to the reference clock is recorded by a TDC.\\
For photon detection, a single-photon detector as described in Ref.~\onlinecite{Hans2018} is used. The photons are passing through an MgF$_2$ window coated with a CsTe layer acting as a photocathode which converts photons into photoelectrons and allows the detection of photons with wavelengths in the range of about $120\,\text{nm}$ to $300\,\text{nm}$ ($4.1\,\text{eV}$ to $10.3\,\text{eV}$).
The electrons are amplified by an MCP chevron stack and the resulting electron cloud hits a delay line type position-sensitive anode \cite{Jagutzki2002,Hans2018}.
The drop of the high voltage at the front MCP is measured using a capacitive coupling and used as the time signal.
While the capability of position sensitive detection is not used for the exemplary measurements presented in this work, future experiments can incorporate these additional information.
While the design of the magnetic bottle provides a relative solid angle $\Omega_{\text{rel},\text{el}}$ close to unity, the relative solid angle $\Omega_{\text{rel},\text{ph}}$ of the photon detection can be estimated to $\Omega_{\text{rel},\text{ph}}\approx\frac{1.26\cdot10^{-3}\,\text{m}^2}{4\pi D^2}$  using Eq. \ref{photodet_eq} with an active area diameter of $40\,\text{mm}$ of the used detector.
In the following, we illustrate two specific optics configurations in detail, which increase $\Omega_{\text{rel},\text{ph}}$  in order to enable electron-photon coincidence experiments within a reasonable data acquisition time.

\section{\label{sec3}Optics design and applications}
\subsection{\label{sec_maxsolid}Configuration optimized for efficiency}
\begin{figure}
\includegraphics{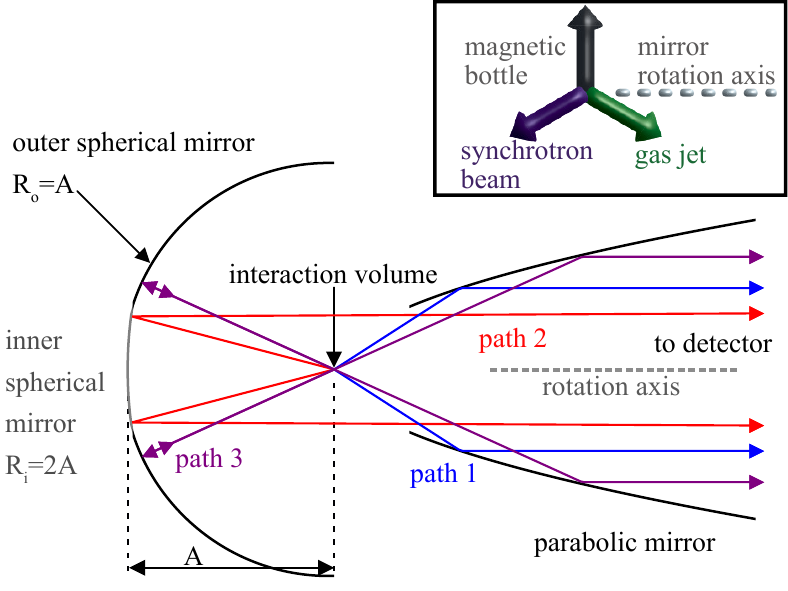}
\caption{\label{fig_hollowmirror}Sketch of the basic rotational symmetric mirror system. The rotation axis is indicated by a grey dashed line.
In the direction of the detector, a parabolic mirror guides the photons onto its active area (path 1, blue).
On the opposite side of the detector, two spherical mirrors reflect photons towards it.
The radius of the inner spherical mirror $R_i$ is twice its distance to the interaction volume $A$ and photons are reflected in a collimated beam to the detector.
The radius of the outer mirror $R_o$ is equal to the distance to the interaction volume $A$ and therefore reflects the photons back into the interaction volume (path 3, violet). From here on, the path coincides with photons of path 1.}
\end{figure}

For this approach, the photon detector is attached to a chamber designed for electron coincidence spectroscopy of gaseous and cluster jets similar as in Ref.~\onlinecite{Hans2016}.
The distance between the active detector surface and the interaction volume is $365\,\text{mm}$, resulting in a relative solid angle (without optics) of $\Omega_{\text{rel},\text{ph}}=0.075\,\%$.
A mirror system for photon detection was designed as illustrated in Figure \ref{fig_hollowmirror} to maximize the solid angle.
It surrounds the complete interaction volume with apertures specifically designed for the used magnetic bottle spectrometer, target jet, and exciting-photon beam.\\
The mirror surfaces are made of polished aluminum to ensure a high reflectance for photons within the sensitivity range of the detector.
The system consists of a combination of a parabolic and two spherical mirrors which guide photons from the interaction volume towards the detector as shown by exemplary ray trajectories in Figure \ref{fig_hollowmirror}:
First, the paraboloid facing the detector parallelizes all photons emitted towards this hemisphere (ray path 1, blue).
Second, the inner spherical mirror possesses a radius such that all photons hitting this mirror from the interaction volume are reflected and also parallelized towards the detector (ray path 2, red). The size of the inner spherical mirror’s area is equal to the entrance width of the parabolic part.
Third, the outer spherical mirror opposite to the detector possesses a radius equal to the distance to the interaction volume, resulting in a reflection back into the interaction volume (ray path 3, violet).
Those rays are then parallelized by the parabolic mirror.

This configuration was tested in an experiment at the \mbox{U49-2 PGM1} beamline (BESSY II, HZB)  \cite{Kachel2016}.
The exciting-photon energy was set to $90\,\text{eV}$.
Using a $50\,\mu\text{m}$ exit slit of the beamline monochromator, the resulting photon beam with a bandwidth of $9\,\text{meV}$ was crossed with a He gas jet.\\
For these experimental conditions, the dominant process is the photoionization of a single 1s electron, \mbox{i.e. $\text{He}(1\text{s}^2)+h\nu\rightarrow\text{He}^+(1\text{s}^1)+\text{e}^-$}.
\begin{figure}[t!]
\includegraphics{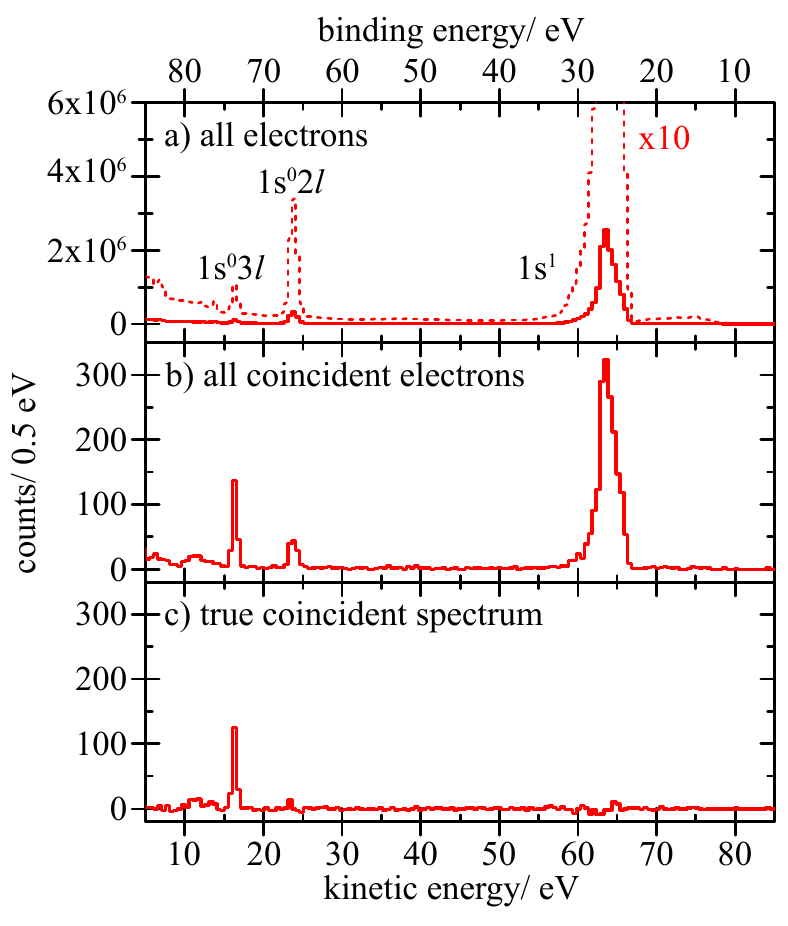}
\caption{\label{fig_He}Electron spectrum of a He gas jet after irradiation with a photon energy of 90 eV.
a) Total electron spectrum (red solid) and its 10-fold magnification (red dashed).
b) All recorded one electron - one photon coincidences (true + accidental).
c) True coincident electron signal after subtraction of accidental coincidences by the method described in section \ref{sec2}.}
\end{figure}
With a comparably low cross section the second electron can be additionally promoted into an excited state during the photoionization process, \mbox{$\text{He}(1\text{s}^2)+h\nu\rightarrow\text{He}^+(1\text{s}^0n\text{p})+\text{e}^-$}, which results in the appearance of so-called satellite lines in the photoelectron spectrum \cite{Heimann1986}.
All satellite states subsequently decay by photon emission, but only the $3\text{p}\rightarrow2\text{s}$ transition with about $7.6\,\text{eV}$ \cite{Kramida2018} transition energy is within the sensitivity range of the employed detector.
The cross section of the $n=3$ satellite is $1.5(2)\,\%$ ,compared to the single 1s ionization \cite{Lindle1987}.
Of course, the 3p electron can also decay to the 1s level with a branching ratio of $\frac{3\text{p}\rightarrow2\text{s}}{3\text{p}\rightarrow1\text{s}}= 0.112$.
If the cross sections for all other processes are neglected, about $0.17\,\%$ of the detected electrons should be in coincidence with a photon in the sensitivity range of the detector.

The $4\text{p}\rightarrow2\text{s}$ transition from the $n=4$ satellite at $10.2\,\text{eV}$ above the ground state lies at the edge of the detector sensitivity and combined with the reduced cross section of the this satellite should lead to a negligible intensity compared to the $n=3$ case.\\ 
The total non-coincident photoelectron spectrum of He is shown in Figure \ref{fig_He}a. The 1s photoelectron line is the most prominent feature followed by the $n=2$ satellite and the suggested appearance of the $n=3$ satellite. In a magnified presentation, the satellite lines up to $n=5$ can be identified (not shown). The energy axis was calibrated using the corresponding energies from Ref.~\onlinecite{Kikas1996}.
If only the coincident electrons are taken into account, the $n=3$ satellite increases in intensity relative to the other features as shown in Figure \ref{fig_He}b.
The elimination of accidental coincidences as described in Section \ref{sec2} yields the (true) photon-coincident electron spectrum shown in Figure \ref{fig_He}c.
Here, only the $n=3$ satellite remains as the radiative $n=3 \rightarrow n=2$ transition is within the sensitivity range of the used detector.
The count rate of true coincidences in this measurement was approximately $0.5\,\text{Hz}$ with a total electron count rate of approximately $58\,\text{kHz}$.
\begin{figure}[t!]
\includegraphics{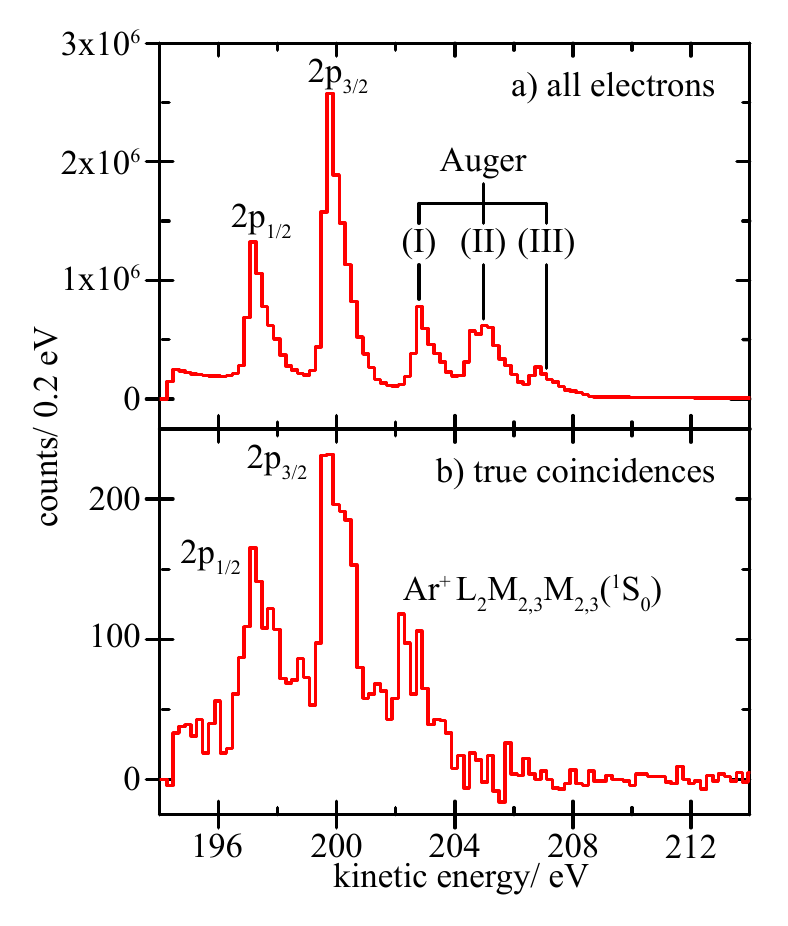}
\caption{\label{fig_Ar}Ar electron spectrum after irradiation with a photon energy of $449\,\text{eV}$.
a) Total electron spectrum, showing the two $2\text{p}$ photoelectron lines and the four Ar Auger lines with the highest kinetic energies (Assignment according to Ref. \onlinecite{McGuire1975}):
(I) \mbox{$\text{Ar}^{+}\,\text{L}_3\text{M}_{2,3}\text{M}_{2,3}(^1\text{D}_{2})$}.
(II) \mbox{$\text{Ar}^{+}\,\text{L}_{3}\text{M}_{2,3}\text{M}_{2,3}(^3\text{P}_{0,1,2})$} and \mbox{$\text{Ar}^{+}\,\text{L}_{2}\text{M}_{2,3}\text{M}_{2,3}(^1\text{D}_{2})$} (not resolved).
(III) \mbox{$\text{Ar}^{+}\,\text{L}_2\text{M}_{2,3}\text{M}_{2,3}(^3\text{P}_{0,1,2})$}. 
b) True coincident electron spectrum. Details are discussed in the text.}
\end{figure}
For an estimate of the effective solid angle of photon detection achieved with this configuration, the ratio of the number of coincidence events to the total intensity of the $n=3$ satellite in the total electron spectrum, which is about $0.0015$, may be used.
This ratio has to be normalized by the $\frac{3\text{p}\rightarrow2\text{s}}{3\text{p}\rightarrow1\text{s}}$ branching ratio and the quantum efficiency of the photon detector.
Since the exact quantum efficiency in the spectral range of the $3\text{p}\rightarrow 2\text{s}$ transition in He II ($165\,\text{nm}$) is not known, for a conservative estimation of the lower limit we use the peak quantum efficiency of $0.255$ at $254\,\text{nm}$\cite{Photek2010}, resulting in an effective solid angle in the order of $\geq5\,\%$.
The geometrical solid angle of the mirror assembly is about $41\,\%$.
We assign the deviation to imperfect reflection of the mirror and variations in the quantum efficiency of the detector. Nevertheless, this conservative estimate results in an increase of the solid angle by a factor of about $70$ compared to the case without optics.\\
As a second example, atomic Ar was photoionized with an exciting-photon energy of $449\,\text{eV}$.
At this photon energy, the 2p photoelectrons and the Ar$^+$ LMM Auger electrons have similar kinetic energies and can be resolved simultaneously by applying a retardation voltage to the drift tube of the magnetic bottle electron spectrometer.
While Auger final states of the form $\text{Ar}^{2+}(3p^{-2})$ cannot decay further, some of the Ar$^+$ LMM Auger channels will end in radiative satellite states of the configuration $\text{Ar}^{2+}(3p^{-3}nl)$.
In Figure \ref{fig_Ar}a, the electron spectrum composed of the 2p$_{1/2}$ and 2p$_{3/2}$ fine structure components and the four Auger channels of highest kinetic energies, corresponding to the $\text{Ar}^{+}\,\text{L}_2\text{M}_{2,3}\text{M}_{2,3}$ and $\text{Ar}^{+}\,\text{L}_3\text{M}_{2,3}\text{M}_{2,3} (^3\text{P}_{0,1,2}\;\text{and}\; ^1\text{D}_2)$ final states, is shown \cite{McGuire1975}.
While the latter Auger electrons should not be accompanied by photon emission, the photoelectrons are, because some 2p vacancies lead to radiative Auger final states (of which the corresponding Auger electrons are not within the detected range).
This is indeed what is observed in Figure \ref{fig_Ar}b, which shows the true photon-coincident electron spectrum.
Surprisingly, one Auger channel is also present at about $203.2\,\text{eV}$ kinetic energy.
We suggest that this weak channel corresponds to radiative decay of the $\text{Ar}^{+}\,\text{L}_2\text{M}_{2,3}\text{M}_{2,3}(^1\text{S}_0)$ Auger final state\cite{McGuire1975} to the $\text{Ar}^{+}\,\text{L}_2\text{M}_{2,3}\text{M}_{2,3}(^3\text{P}_{0,1,2})$ state via magnetic dipole or electronic quadrupole transitions, which are within the sensitivity range of the detector.
\subsection{\label{sec_adapt}Configuration optimized for adaptability}
\begin{figure}
\includegraphics{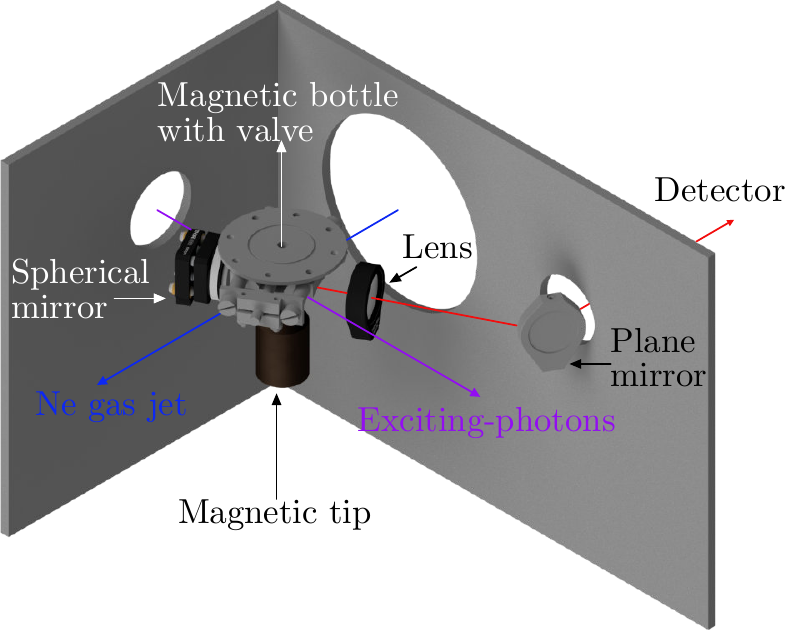}
\caption{\label{fig_adaptset}Isometric view of the adaptable set-up.
A spherical mirror and a plano convex lens are used to increase the solid angle for the photon detection.
The plano convex lens distance and focal length are chosen such that the transmitted photons are parallelized while the plane mirror deflects the photons onto the detector axis.
Therefore, an observation of the emitted photons without direct view is possible.}
\end{figure}

In certain cases, the application of configuration A might be hindered by experimental constraints.
For example, the layout of the vacuum chamber, target source, or electron spectrometer can interfere spatially with the mirror or the direct view towards the interaction volume might be blocked.
Then, the solid angle can still be increased significantly using a combination of flexible optical elements.\\
In the presented experiment, a magnetic bottle electron spectrometer similar to configuration A was used with a different interaction chamber.
However, no port with a direct view of the interaction volume was available for the photon detector as illustrated in Figure \ref{fig_adaptset}.
In addition, a valve at the entrance of the electron spectrometer would intersect with parts in the close vicinity of the interaction volume thus preventing a mirror design as described above.
With these constraints, an assembly of three different optical elements is used for photon guidance.
A combination of an UV enhanced Al coated spherical mirror and a fused silica spherical plano convex lens, positioned at opposite sides of the interaction volume, may achieve a relative solid angle of up to $7.5\,\%$.
Simultaneously, the lens collimates the emitted photons onto a planar mirror, which is used to redirect the photons onto the detector.
The functionality of this configuration was validated in an experiment conducted at the \mbox{UE56/2 PGM1} beamline at BESSY II in Berlin.
Neon atoms were injected effusively into the interaction chamber through a $25\,\mu\text{m}$ nozzle.
The exciting-photon energy was set to $867.1\,\text{eV}$ corresponding to the resonant $1\text{s}^22\text{s}^22\text{p}^6 \rightarrow1\text{s}^12\text{s}^22\text{p}^63\text{p}$ excitation of atomic Ne.
Here, a variety of different de-excitation pathways are possible \cite{Hayaishi1995}.
\begin{figure}
\includegraphics{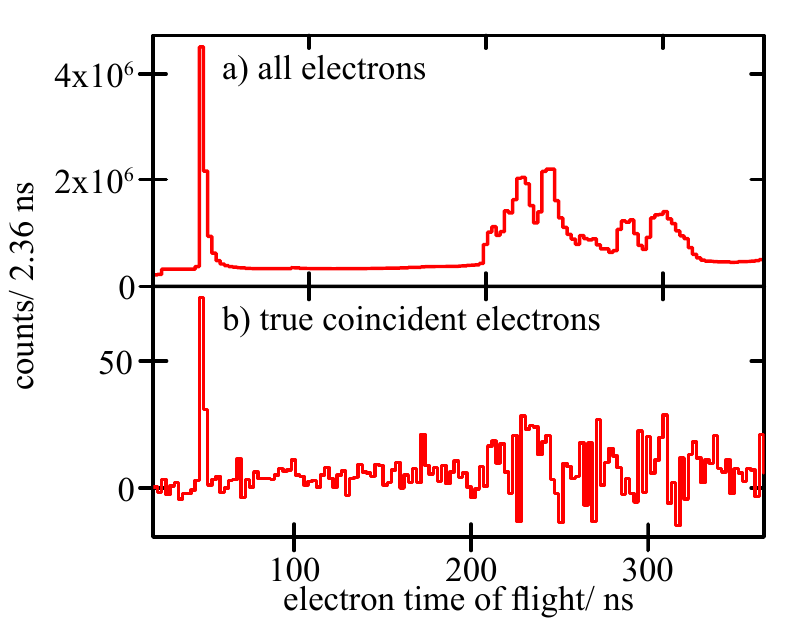}
\caption{\label{fig_Ne}Electron spectra of Ne after irradiation with a photon energy of $867.1\,\text{eV}$.
a) Total electron spectrum.
b) True coincident electron spectrum.
}
\end{figure}
The total electron spectrum in Figure 6a consists of an intense peak at short times of flight, comprised of unresolved Auger and valence electrons.
The slower electrons are the result of further autoionizing Auger final states. However, the only relaxation channels observable by an electron-photon coincidence are spectator Auger final states of the form $1\text{s}^22\text{s}^22\text{p}^4n\text{p}$ with fast Auger electrons included in the fast electron peak.
Consequently, the autoionizing Auger final states vanish in the true coincidence spectrum in Figure 6b.
Here, the true coincidence rate was about $0.01\,\text{Hz}$, compared to a total electron count rate of about $60\,\text{kHz}$.
Despite the low count rate, the experiment yields an interpretable result.
This illustrates that the coincidence measurement involves an extremely efficient noise reduction, allowing to detect signals orders of magnitude weaker than the non-coincident-noise.
\section{\label{sec4}Conclusion}
We have developed two optical assemblies which can be combined with magnetic bottle electron spectrometers to result in a highly efficient set-up for electron-photon coincidence experiments.
While configuration A focuses on the optimization of the solid angle of the photon detection, configuration B is adaptable to different experimental constraints.\\
Both configurations were tested successfully for exemplary physical processes that feature electron-photon coincidences after excitation of atomic noble gas targets with synchrotron radiation.
We demonstrated that this method is capable of identifying obscured physical processes and can circumvent the signal to noise ratio problem of very low count rate experiments.
We envision this method to be capable to unravel energy and charge transfer processes in dense media.
Here, the additional insight of photon detection allows to further characterize ultra-fast phenomena taking place only in such dense media.
Examples where photon emission plays an decisive role are resonant Interatomic Coulombic Decay\cite{Knie2014}, Radiative Charge Transfer\cite{Hans2018a}, or possibly ultra fast proton transfer in liquid samples\cite{Hans2017,Thuermer2013}.
\begin{acknowledgments}
We thank Claudia Kolbeck, Sebastian Malerz, Marvin Pohl, Anne Stephansen (FHI Berlin, Germany) and Clara Saak (Uppsala University, Sweden) for their support during the synchrotron radiaton beamtime. 
We gratefully acknowledge the HZB for the allocation of synchrotron radiation beamtime and the BESSY II staff for their support. This work was funded by the Deutsche Forschungsgemeinschaft (DFG, German Research Foundation) – Projektnummer 328961117 – SFB 1319 ELCH and Research Unit FOR 1789.
\end{acknowledgments}
\bibliography{coinc_bib}
\end{document}